\documentclass[fleqn,10pt]{wlscirep}
\usepackage{amsmath,amssymb,graphicx,color}
\usepackage{cancel}
\usepackage{soul} 
\usepackage{epstopdf}

\begin{document}
\title{Chiral Separation by Flows: The Role of Flow Symmetry and Dimensionality}

\author[1]{Sunghan Roh}
\author[2]{Juyeon Yi}
\author[1,*]{Yong Woon Kim}

\affil[1]{Graduate School of Nanoscience and Technology, Korea Advanced Institute of Science and Technology, Deajeon 34141, Korea}
\affil[2]{Department of Physics, Pusan National University, Busan 46241, Korea}

\begin{abstract}
Separation of enantiomers by flows is a promising chiral resolution method using cost-effective microfluidics.
Notwithstanding a number of experimental and numerical studies, a fundamental understanding still remains elusive, 
and an important question as to whether it is possible to specify common physical properties of flows that induce separation has not been addressed.
Here, we study the separation of rigid chiral objects of an arbitrary shape induced by a linear flow field at low Reynolds numbers.
Based on a symmetry property under parity inversion, 
we show that the rate-of-strain field is essential to drift the objects in opposite directions according to chirality.
From eigenmode analysis, we also derive an analytic expression for the separation conditions 
which shows that the flow field should be quasi-two-dimensional for the precise and efficient resolutions of microscopic enantiomers.
We demonstrate this prediction by Langevin dynamics simulations with hydrodynamic interactions fully implemented. 
Finally, we discuss the practical feasibility of the linear flow analysis, considering separations by a vortex flow or an extensional flow under a confining potential.
\end{abstract}

\maketitle

\section*{Introduction}

An enantiomer is a molecule that cannot be superposed on its own mirror image, and such property is called chirality~\cite{Stalcup}. 
Despite the structural similarity, a pair of enantiomers often exhibits very different biochemical activities due to the chiral nature of living systems~\cite{Ariens}.
It is, therefore, of great importance to separate a racemic mixture by chirality in the pharmaceutical, agricultural, and environmental industries where a number of molecules in use are enantiomers~\cite{Maier}.
The conventional separation techniques such as chromatography and capillary electrophoresis require molecular specific and expensive chiral agents or media. 

Among alternative physical separation methods that do not rely on a chiral selector~\cite{deGennes, Baranova, Schamel}, 
chiral resolution by flows has recently received considerable attention with rapid developments in microfluidics~\cite{Howard, Kim, deGennes, Chen1, Makino, Marcos, Maria, Hermans, Kostur, Meinhardt, Eichhorn, Chen, Watari, Baranova, Schamel, Talkner}.
Since the original suggestions~\cite{Howard, Kim}, chirality-dependent drift has been demonstrated by several experiments using shear or vortex flows~\cite{Chen1, Makino, Marcos, Maria, Hermans}.
While a number of numerical studies~\cite{Kostur, Meinhardt, Eichhorn, Chen, Watari, Talkner} have been done to propose various sorting strategies by assuming particular flow fields, e.g., microfluidic vortices~\cite{Kostur} or asymmetric flows with different slip lengths~\cite{Meinhardt, Eichhorn},
there are few efforts to develop a general theoretical framework for arbitrary linear flow patterns and object shapes.
Hence, important questions, what are the common characteristics, if any, of flows that cause separation?
or what is the role of each flow component in separation?, are still to be answered.

In this work, we address this problem by introducing a theoretical framework to understand the motion of a rigid chiral object of any shape in an arbitrary linear Stokes flow.
The essential flow component for separation is elucidated from a symmetry argument using parity inversion and mirror reflection.
We also show that the velocity gradient tensor has to be nearly singular, i.e., quasi-two-dimensional, to induce the separation of high precision. 
In order to validate our analytic results, we perform Langevin dynamics simulations
that explicitly incorporate hydrodynamic interactions among object elements.
In simulations, the separation precision is quantified by the Jensen-Shannon divergence between the probability distributions of particles of opposite chirality.

\section*{Results}

\subsection*{Model}

Consider a rigid chiral object moving through a viscous incompressible fluid.
Let $\vec{r}$ be the position of any point $O$ fixed to the particle, say, the center of mass, with respect to a reference frame.
At low Reynolds number (i.e., in the inertialess regime), the equations of motion are obtained from the Stokes equation for 
translational velocity $\vec{v}$ of that point and angular velocity $\vec{\omega}$ of the object, in relative to the flow~\cite{Sangtae, Brenner}: 
\begin{equation}\label{Eq:EquationOfMotion}
 \left(
    \begin{array}{c}
     \vec{ v}-\vec{ U}\\ \vec{ \omega}-\vec{ \Omega}

    \end{array}
\right)
= \left(
    \begin{array}{cc}
    {\boldsymbol \mu}^{\textrm{tt}} & {\boldsymbol \mu}^{\textrm{tr}} \\
    {\boldsymbol \mu}^{\textrm{rt}} & {\boldsymbol \mu}^{\textrm{rr}}
    \end{array}
\right)
\left(
    \begin{array}{c}
    {\boldsymbol \zeta}^{\textrm{te}} \\ {\boldsymbol \zeta}^{\textrm{re}}
    \end{array}
\right) : {\bf E}
+\left(
    \begin{array}{cc}
    \vec{\xi^{\textrm{t}}} \\
    \vec{\xi^{\textrm{r}}}
    \end{array}
\right)
\end{equation}
where $\vec{U} \equiv \vec{U} (\vec{r})$ denotes the ambient fluid velocity at $\vec{r}$,
and $\vec{\Omega}=\frac{1}{2}\nabla \times \vec{U} (\vec{r})$ is half the fluid vorticity.
Here the ambient flow is approximated as a linear Stokes flow $\vec{U}({\vec r})= \vec{U}(0)+{\bf J}{\vec r}$ 
with a constant velocity gradient tensor ${\bf J}=\nabla \vec{U}$~\cite{Sangtae}.
The rate-of-strain field ${\bf E}$ is the symmetric part of ${\bf J}$, 
given as ${\bf E}=({\bf J}+{\bf J}^{\textrm T})/2$ with the transpose operator $\textrm T$. 
The mobility tensors, ${\boldsymbol \mu}$'s, are related to the strengths of thermal noises, $\vec{\xi}$'s,  through the fluctuation dissipation theorem
$\left\langle \xi_i^m \xi_j^n \right\rangle = 2  k_B T {\boldsymbol \mu}^{m n}_{ij}$,
with $m, n = \textrm{t}, \textrm{r}$ and $k_B T$ being the thermal energy at temperature $T$. 
Here the superscripts, $\textrm{t}$ and $\textrm{r}$, stand for `translational' and `rotational', respectively;
${\boldsymbol \mu}$'s couple two degrees of freedom corresponding to its superscripts.  
The third-rank resistance tensors , ${\boldsymbol \zeta}$'s, determine hydrodynamic friction force
and couple the translational and rotational motions of the object with the rate-of-strain field 
(for which we assign `$\textrm{e}$' as a superscript symbol, and see also index notation of Carrasco et. al~\cite{Garcia}.)

The tensor product in Eq.~(\ref{Eq:EquationOfMotion}) is defined as $ \left( {\boldsymbol \zeta} : {\bf E} \right)_i \equiv \sum_{j,k} {\boldsymbol \zeta}_{ijk} {\bf E}_{jk}$. 
Formal complexity arising from the no-slip condition on the object surface lies in calculating ${\boldsymbol \mu}$ and ${\boldsymbol \zeta}$
that depend on the position vectors of surface elements relative to $O$.
In calculating ${\boldsymbol \mu}$ and ${\boldsymbol \zeta}$, the no-slip condition on object surface is imposed, and they depend on the position vectors
of the surface elements relative to a certain origin $O$.
The mobility and resistance tensors are thus independent of the choice of the origin,
and they are functions only of the geometry of the particle such as object orientation ${\hat \varphi}$ and handedness $\alpha$.
Therefore, the chirality-dependent drift can be induced only by the rate-of-strain field ${\bf E}$.  
We write it in shorthand notation,
\begin{equation}\label{driftvel}
{\vec v}_{\mathbf{E}}^\alpha ({\hat \varphi})\equiv  
{\boldsymbol \mu}^{{\textrm{tt}}} ( {\boldsymbol \zeta}^{{\textrm{te}}} : {\bf E} ) +
{\boldsymbol \mu}^{{\textrm{tr}}} ( {\boldsymbol \zeta}^{{\textrm{ re}}} : {\bf E} ) ,
\end{equation}
with chirality index $\alpha=R$ or $L$ and refer to it as {\it drift velocity}.

\begin{figure}[t]
	\center
	\includegraphics[width=0.9\linewidth]{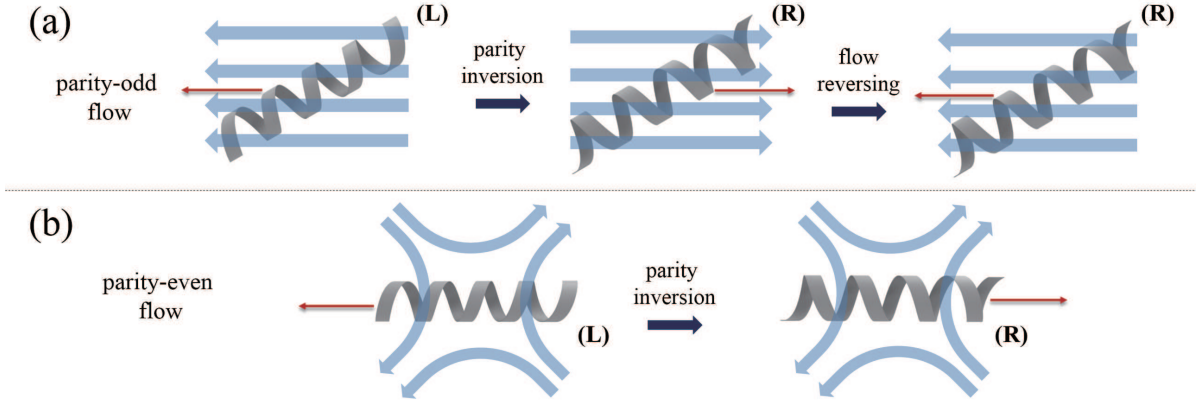}
	\caption{Schematic figure illustrating motions of a chiral object and its chiral partner under (a) parity-odd and (b) parity-even flows. 
	The chiral objects are represented by the helices, the flow fields by blue arrows, and red arrows indicate the translational motions of objects induced by the flow.
	}
	\label{Fig:Parity}
\end{figure}

\subsection*{Symmetry under parity inversion}

Using a symmetry property under parity-inversion operation, we first show that the chiral separation (objects move in opposite directions according to handedness indeed) occurs by the parity-even rate-of-strain field.
In particular, this is proven even with considering the rotational motions of object orientations.
Using parity-inversion ($\mathbb{P}$) operation about the point $O$, a linear flow is decomposed into two parts;
one is a parity-odd flow whose direction is changed by $\mathbb{P}$ operation, and the other is a parity-even flow that remains invariant under $\mathbb{P}$ operation.

In Fig.~1, translation motions of a chiral molecule and its chiral partner under parity-odd and parity-even flows are illustrated, respectively. Shown on the left panel of Fig.~1(a) is a left-handed ($L$) object in a parity-odd flow, which is supposed to drift to the left by the flow. Applying $\mathbb{P}$ operation on this system, the flow direction is inverted, and the object handedness as well as the drift direction are reversed (middle panel). 
If the flow field becomes again inverted (right panel), the original flow field is recovered and the right-handed ($R$) object then moves to the left
because of the linearity of relations between the object velocity and the flow fields.
One clearly sees that the drift motions of the chiral pair are identical under the parity-odd flow, leading to no chirality resolutions.
On the contrary, if $\mathbb{P}$ operation is applied on an object of $L$ in a parity-even flow (Fig.~1(b)), the object handedness is converted into $R$ and the drift direction is changed as well, while the flow itself remains invariant. 
This enables us to have the opposite motions of the chiral pair under the same ambient flow and
thus demonstrates that the parity-even flow field is essential for chiral separation. 

According to Eq.~(\ref{Eq:EquationOfMotion}), the translational velocity $\vec{v}$ of an object includes two flow components $\vec{U}$ and $\mathbf{E}$. It is obvious that $\vec{U}$ is parity-odd while $\mathbf{E}$ is parity-even as $\mathbf{J}$ remains invariant under $\mathbb{P}$ operation. Therefore, from the analysis above, we find that the drift velocity of a left-handed object is opposite to its chiral partner under a given rate-of-strain field:
\begin{equation}
	\vec{v}_{\mathbf{E}}^L \left( \hat{\varphi} \right) = - \vec{v}_{\mathbf{E}}^R \left( \hat{\varphi}' \right),
\end{equation}
where $\hat{\varphi}'$ denotes parity-inverted orientation of $\hat{\varphi}$.
It is to be noted that the angular motion remains invariant under $\mathbb{P}$ operation, and thus the probability distributions of orientation are symmetric, $\Phi^L({\hat {\varphi}}) = \Phi^R({\hat {\varphi}'})$, at any time $t$ (see the Supplemental Information for details).  
Accordingly, the orientation-averaged drift velocity of the left-handed object is given by
\begin{eqnarray}
\label{Eq:v_parity}
\langle {\vec v}^{L}_{\mathbf E}\rangle_{\Phi}
&=& \int d{\hat {\varphi}} ~{\vec v}^{L}_{{\mathbf E}}(\hat{\varphi}) ~\Phi^L({\hat {\varphi}}) \\ \nonumber
&=& -\int d{\hat {\varphi}}' ~{\vec v}^{R}_{{\mathbf E}}(\hat{\varphi}') ~\Phi^R({\hat {\varphi}}') 
= - \langle {\vec v}^{R}_{\mathbf E}\rangle_{\Phi} ,
\end{eqnarray}
where the negative sign of the average drift velocity implies a possibility of chiral separation. 

In a similar way, the separation direction can be specified if a linear flow field has a mirror-reflection symmetry.
Shear flow, $\vec{U}({\vec r}) = \dot{\gamma}y{\hat x}$ with shear rate $\dot{\gamma}$, belongs to the case, if choosing the $xy$-plane as a mirror-symmetry plane. 
Regard a relation, 
\begin{equation}\label{mirror}
 (v^L_{{\mathbf E},\parallel }(\hat{\varphi}), v^L_{{\mathbf E}, \perp}(\hat {\varphi}))
=(v^{R}_{{\mathbf E}'',\parallel }(\hat{\varphi}''), -v^{R}_{{\mathbf E}'', \perp}(\hat {\varphi}'')),
\end{equation}
where the double prime indicates quantities transformed by mirror-reflection ($\mathbb{M}$) operation. 
Here, $v^{\alpha}_{\mathbf{E},\parallel}$ and $v^{\alpha}_{\mathbf{E},\perp}$ are, respectively, the parallel and perpendicular components of a drift velocity 
$\vec{v}^{\alpha}_{\mathbf{E}}$ to a chosen mirror-reflection plane.
For a mirror-symmetric flow field (${\mathbf E}''={\mathbf E}$),
taking average over object orientations for $\Phi^L({\hat {\varphi}}) = \Phi^R(\hat {\varphi}'')$, one has a relation, 
\begin{eqnarray}
\label{Eq:v_mirr}
(\langle  v_{{\mathbf E},\parallel}^{L}\rangle_{\Phi}, \langle v_{{\mathbf E},\perp}^{L}\rangle_{\Phi} ) = 
(\langle v_{{\mathbf E},\parallel}^{R}\rangle_{\Phi}, -\langle v_{{\mathbf E},\perp}^{R}\rangle_{\Phi} ).
\end{eqnarray}
Combining with Eq.~(\ref{Eq:v_parity}), one finds that $\langle v^{\alpha}_{{\mathbf E},\parallel}\rangle_{\Phi}$ =0, and 
the separation of objects with opposite chirality can occur along the direction perpendicular to the mirror-symmetry plane, as indeed in the case of the shear flow. 
We again note that Eqs.~(\ref{Eq:v_parity}) and (\ref{Eq:v_mirr}) are obtained
through the ensemble averages over time-dependent probability distributions of orientations.

\subsection*{Separation criterion}

The nonvanishing average drift velocity $\langle {\vec v}^{\alpha}_{\mathbf E}\rangle_{\Phi}$ is, though essential, only a necessary condition.
For practical realizations of chiral separations, the position dependence of $\vec{U}$ should be analyzed. In other words,
the chirality-independent drift by $\vec{U}(\vec{r})$ should not dominate the chirality-dependent drift $\vec{v}_{\mathbf E}^\alpha$ by $\mathbf E$.
Furthermore, the positional distributions of particles should evolve by $\vec{v}_{\mathbf E}^\alpha$ to be well separated on macroscopic scales, at least, 
larger than experimental resolutions of microfluidic devices. 
Below we argue that this is achieved with quasi-two-dimensional flow fields.

\subsubsection*{eigenmode analysis}

It is more intuitive to examine the equation for $\vec{v}$ for the case of diagonalizable $\mathbf{J}$
(for a non-diagonalizable case the analysis can be performed in a similar way, leading to the qualitatively same conclusions; see the Supplemental Information for details).
The equation for $\vec{v}$ in Eq.~(\ref{Eq:EquationOfMotion}) can be written in the diagonalizing basis of $\mathbf{J}$,
and one of its components along the direction of the $q$-th eigenvector of $\mathbf{J}$ reads as,
\begin{equation}\label{eom}
\dot{r}^\alpha_{\lambda_{q} }= \lambda_{q} r^\alpha_{\lambda_{q}} + v^\alpha_{{\mathbf E}, \lambda_{q}} + \xi_{\lambda_{q}}^{\textrm{t}},
\end{equation}
where $\lambda_{q}$ is the $q$-th eigenvalue of ${\bf J}$, and quantities with the subscript $\lambda_{q}$ symbolize the $q$-th component of transformed vectors $({\bf S} {\vec X})_{q} = X_{\lambda_{q}}$ with
$({\bf S}{\bf J}{\bf S}^{-1})_{q,q'}=\lambda_{q}\delta_{q,q'}$ and $X=r,v_{\mathbf E},\xi^{\textrm{t}}$. 
For notational simplicity, we shall drop the eigenvalue index $q$ and chirality index $\alpha$ hereafter.
Without loss of generality, we have set $\vec{U}(0)=0$.

The formal solution of Eq.~(\ref{eom}) is given by
\begin{equation}\label{sol1}
r_{\lambda}(t) = e^{\lambda t} r_{\lambda}(0)+ X_{{\mathbf E},\lambda}(t) + \Xi_{\lambda}(t).
\end{equation}
Here we define the displacements resulting from the drift motion, $X_{{\mathbf E},\lambda}(t)=\int_{0}^{t} dt' e^{\lambda(t-t')} v_{{\mathbf E},\lambda}(t') $, and from the thermal noise $\Xi_{\lambda}(t)= \int_{0}^{t} dt' e^{\lambda(t-t')} \xi_\lambda^{\textrm{t}}(t')$. Taking average over the thermal noises and initial positions, we obtain the average displacement 
\begin{equation}\label{dissol}
\left\langle r_\lambda (t) \right\rangle = \langle X_{{\mathbf E},\lambda}(t)\rangle
\end{equation}
with the average of initial positions located at origin. 
Mean separation distance between particles having opposite
chirality is given by ${\cal D}(t) = | \langle r^{R}_\lambda (t) \rangle - \langle r^{L}_\lambda (t) \rangle |$, and separation precision
can be quantified by ${\cal D}(t)/\Sigma(t)$ with $\Sigma^{2}(t) \equiv \langle r^{2}_{\lambda}(t)\rangle - \langle r_{\lambda}(t)\rangle^{2}$. The chiral separation is achieved if the quantifiers satisfy following relations:
\begin{eqnarray} \label{Eq:conditions}
{\cal D}(t)/\Sigma(t) \gg 1 ~~~ \mathrm{and} ~~~ {\cal D}(t)/\sigma_0 \gg 1~.
\end{eqnarray}
The first condition requires that the mean separation distance ${\cal D}(t)$ should be much larger than the dispersion $\Sigma(t)$ for efficient separation. 
The second condition states that the separation distance ${\cal D}(t)$ ought to be much larger than the initial width of distribution 
$\sigma_{0}$ which roughly amounts to experimental resolution such as the distance between recovery outlets of devices for the separated particles.

It is clear that an exact analysis of the separation conditions (\ref{Eq:conditions}) requires to obtain the precise form of $v_{{\mathbf E},\lambda}(t)$ 
which is highly nonlinear and has no known solution to the best of our knowledge. Notwithstanding the difficulty, one can still put forward a reasonable analysis:
Suppose that there exists a finite maximum drift velocity $v_m$ of $v_{{\mathbf E},\lambda}(t)$, viz. $v_{{\mathbf E},\lambda}(t) \leq v_{m}$,
which is reasonable because the drift velocity as given in Eq.~(\ref{driftvel}) is determined by the product of the bounded quantities.
It is also assumed that up to the leading order, the maximum drift velocity $v_m$ is proportional to the magnitude of flow field given as $|{\bf J}| \sim \sqrt{\sum_i^3 \lambda_i^2}$,
and for a fixed $|{\bf J}|$, only weakly depends on an individual $\lambda_i$.
This assumption may be supported by an observation that as increasing the flow gradient tensor ${\bf J}$ by a factor of a certain constant, 
the resulting drift velocity will be increased proportionally.
Lacking in mathematical rigour, this ansatz is effective to extract an essential flow property to induce chiral separation, as evidenced later in our numerical simulations.
Since $\int \,d x ~ f(x) \leq \int \,d x ~ g(x)$ for $f(x) \leq g(x)$, the maximum value of the mean separation is determined by $r_{\lambda}(t)$ in Eq.~(\ref{sol1}) with letting $v_{{\mathbf E},\lambda}(t) =v_{m}$ as
\begin{eqnarray}\label{max}
d(t)\equiv \mbox{max}[{\cal D}(t)]&=&2\left\langle [r_\lambda (t)]_{ v_{{\mathbf E},\lambda}(t)=v_{m}}\right\rangle~ \\ \label{constvel1}
&=&2(v_{m}/\lambda) (e^{\lambda t}-1) ,
\end{eqnarray}
On the other hand, the positional dispersion can be written as  
\begin{eqnarray}\nonumber
\Sigma^{2}(t)&=&\sigma^{2}(t)+\langle X^{2}_{{\mathbf E},\lambda}(t)\rangle - \langle X_{{\mathbf E},\lambda}(t)\rangle^{2}, \\ \label{constvel2}
\sigma^{2}(t)&=&  \sigma_{0}^{2} e^{2\lambda t}+ (D/ 2\lambda) \left( e^{2\lambda t} - 1\right),
\end{eqnarray}
where $\sigma(t)$ determines the dispersion when $v_{{\mathbf E},\lambda}(t)$ is purely deterministic to annihilate fluctuations 
of $X_{{\mathbf E},\lambda}(t)$. Obviously, an inequality, $\Sigma(t) \geq \sigma(t)$, follows, which together with Eq.~(\ref{max}) leads to ${\cal D}(t)/\Sigma(t) \leq d(t)/\sigma(t)$
and ${\cal D}(t)/\sigma_{0} \leq d(t)/\sigma_{0}$. Eq.~(\ref{Eq:conditions}) then constitutes a necessary condition for chiral separation,
\begin{eqnarray} \label{Eq:conditions2}
d(t)/\sigma(t) \gg 1 ~~~ \mathrm{and} ~~~ d(t)/\sigma_0 \gg 1~,
\end{eqnarray}
which is given in a greatly simplified form to allow an analytic approach and helps to extract essential flow factor inducing chiral separation. 
It should also be mentioned that in high P\'{e}clet number regime the drift velocity as a function of object orientation remains roughly constant,
as suggested by Marcos {\it et. al.}~\cite{Marcos}. For the case, $v_{m}$ can be interpreted as the constant drift velocity and hence,
conditions in Eq.~(\ref{Eq:conditions2}) are equivalent to Eq.~(\ref{Eq:conditions}).  In the following, we proceed our analysis of the separation conditions,
(\ref{Eq:conditions2}) with $d(t)$ and $\sigma(t)$ determined by Eq.~(\ref{constvel1}) and Eq.~(\ref{constvel2}), respectively.

\subsubsection*{short time behavior}

\begin{figure}
	\center
	\includegraphics[width=1.0\linewidth]{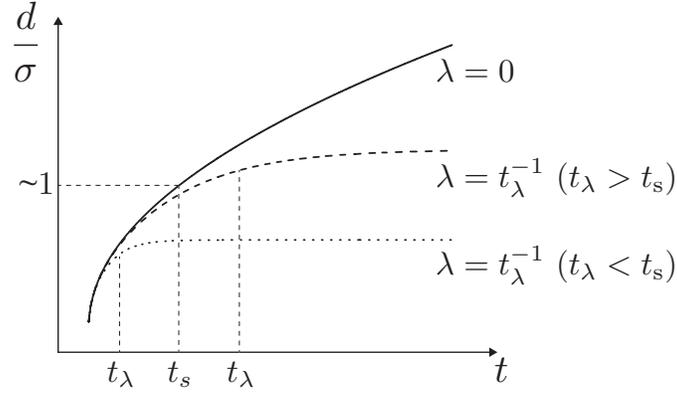}
	\caption{Separation precision, $d/\sigma$, as a function of time $t$ at three different eigenvalues: $\lambda=0$ (solid line), small $\lambda$ with $t_\lambda = |\lambda|^{-1} \gg t_s$
	(dashed line), and large $\lambda$ with $t_\lambda = |\lambda|^{-1} \ll t_s$ (dotted line).
	Here, $t_s$ indicates the separation time scale at which $d/\sigma \approx 1$.
	For a vanishing eigenvalue $\lambda$ (i.e., a singular velocity gradient tensor), $d/\sigma$ monotonically increases with $t$, while
	for a non-zero eigenvalue, $d/\sigma$ is saturated to a constant value after the saturation time scale, $t_\lambda$. 
	}
	\label{Fig:schematic}
\end{figure}

We examine the behavior of the separation precision for $ t \ll t_\lambda \equiv |\lambda|^{-1}$, where $t_\lambda$ is the saturation time scale at which  
the separation precision with a non-zero $\lambda$ approaches a constant value. 
In Fig.~\ref{Fig:schematic}, we illustrate the characteristic behaviors of $d/\sigma$ depending on $\lambda$. 
For $\lambda =0$, $d/\sigma$ grows unboundedly with time $t$, while for $\lambda\neq 0$, it approaches a constant value after a time $t_\lambda$.
That is, in the case of $\lambda \neq 0$, the feasibility of $d/\sigma \gg 1$  depends on the system parameters such as initial dispersion and diffusion constant, 
even at large $t$. 
For $t \ll t_\lambda $, $d$ is monotonically increasing function of time, and irrespective of the sign of the eigenvalue $\lambda$, $d/\sigma$ is simplified as
\begin{equation}
\label{Eq:shorttimes}
\frac{d(t)}{\sigma(t)} \sim \frac{t}{\sqrt{t_{\sigma}^{2}+t_{D}t}}, 
\end{equation}
where $t_\sigma \equiv \sigma_0 /v_m$ is the time required for an object to travel a distance of initial positional dispersion $\sigma_0$ by the  drift velocity $v_m$,
and $t_D \equiv  D/v_m^2$ is the time scale at which the traveling distance by the drift motion is comparable to the diffusion length ($v_m t_D \sim \sqrt{D t_D}$).
From the separation conditions, Eq.~(\ref{Eq:conditions2}), and the equation of the separation precision, Eq.~(\ref{Eq:shorttimes}), one can readily define the separation time scale $t_s$ at which both $d/\sigma$ and $d/\sigma_0$ become of the order of unity (see Fig.~\ref{Fig:schematic}) as
\begin{equation}
t_s \equiv \mbox{max} \left( t_\sigma, t_D \right) ,
\end{equation}
and therefore, an efficient and precise chiral resolution (i.e., $d/\sigma \gg 1$ and $d/\sigma_0 \gg 1$) occurs when
\begin{equation}
\label{Eq:condition1}
t_\lambda \gg t \gg t_s~.
\end{equation}
This constitutes the condition of macroscopic chiral separation for short times, i.e., $t\ll t_\lambda$.
 
\subsubsection*{long time behavior}

In the long time regime of $t \gg t_\lambda$, if the eigenvalue is not zero ($\lambda \neq 0$), $d/\sigma$ is saturated to a certain value as
\begin{equation}
\label{Eq:longtimes}
\frac{d}{\sigma} \sim \left\{ 
\begin{array} {cc}
\displaystyle{ \frac{t_\lambda }{\sqrt{t_{\sigma}^{2}+ t_{D} t_{\lambda}}} }& (\lambda > 0) \\
\displaystyle{ \sqrt{ \frac{t_\lambda}{ t_D}}} & (\lambda < 0) .
\end{array}
\right.
\end{equation}
When $\lambda > 0$, it is found from Eq.~({\ref{Eq:longtimes}) that the condition of macroscopic separation ($d/\sigma \gg 1$) is translated into
\begin{equation}
\label{Eq:positive}
t_\lambda \gg t_s.
\end{equation}
This can also be intuitively understood from Fig.~\ref{Fig:schematic}, i.e., only when $t_\lambda \gg t_s$, the saturated value of $d/\sigma$ can be much greater than one.
Then, the condition of $d/\sigma_0 \gg 1$ is always satisfied since $d/\sigma_0 \sim (t_\lambda/t_\sigma) (e^{t/t_\lambda} -1 )$. 
On the other hand, when $\lambda < 0$, 
the first condition of Eq.~(\ref{Eq:conditions2}) leads to $t_\lambda \gg t_D$, and
the second condition does to $t_\lambda \gg t_\sigma$.

Remarkably, we find that in both of short-time and long-time regime, the macroscopic chiral separations are dictated by a single criterion, irrespectively of sign of $\lambda$:
\begin{equation}
\label{Eq:criterion}
t_{\lambda} \gg t_{s}~~\mbox {or}~~\lambda \ll \frac{1}{t_s} = \mbox{min} \left(\frac{1}{ t_\sigma}, \frac{1}{t_D} \right) ,
\end{equation}
which means that the eigenvalue of the velocity gradient tensor should be significantly smaller than the inverse of the separation time scale.
We note that the eigenvalue index $q$ has been omitted for notation simplicity. 
Therefore, Eq.~(\ref{Eq:criterion}) should be interpreted as conditions required to be satisfied by respective $\lambda_q$ in order to obtain the separation along the corresponding coordinate $r_{\lambda_q}$. In other words, the chiral separation may occur if at least one of the eigenvalues satisfies the condition.

\subsubsection*{Dimensional analysis and estimates of parameters}  

Let us now express the separation condition, Eq.~(\ref{Eq:criterion}), in terms of physical parameters such as 
the linear size of the object $\ell$ and magnitude of the flow velocity gradient $V$ defined through $\mathbf{J}$ (for example, see Eqs.~(\ref{Eq:JA}) and (\ref{Eq:JB}) of the next section).
One may define $V$ also as $V\sim\sqrt{\sum_{i,j=1}^{3}J_{ij}^{2}}$, where the proportionality constant is of the order of unity, and it is irrelevant to the present analysis.
We introduce dimensionless parameters  
\begin{equation*}
\delta \equiv \sigma_0/\ell,~~~ \epsilon \equiv \lambda / V,\,\,\, c \equiv v_m/V\ell, 
\end{equation*}
where $\delta$ is initial dispersion $\sigma_0$ in units of $\ell$, and it seems reasonable to assume $\delta \gg 1$
in most cases of practical interest of small particles. $\epsilon$ is the amplitude of $\lambda$ relative to the magnitude of the flow velocity gradient, which can characterize the dimensionality of the flow field; if $\epsilon=0$, the corresponding flow becomes a two-dimensional flow. 
$c$ measures the chiral-dependent drift velocity relative to a variation of ambient flow velocity over $\ell$.
Since $D \sim k_B T /\eta \ell$ with the solvent viscosity $\eta$, $t_\sigma = \sigma_0/v_m \sim \delta/cV$ and $t_D = D/v_m^2 \sim k_B T / c^2 V^2 \eta \ell^3$. 
As a result, the separation condition of Eq.~(\ref{Eq:criterion}) can be written in dimensionless form as
\begin{equation} 
\label{Eq:dim_result}
|\epsilon| \ll \textrm{min} \left(\frac{c}{\delta}, c^2 \, \textrm{Pe} \right)
\end{equation} }
where $\mathrm{Pe} \equiv V /D_r$ with the rotational diffusion constant $D_r \sim k_B T/\eta \ell^3$.

Among the parameters, $c$ affects both of $t_\sigma$ and $t_D$, and its estimation is important for numerical evaluations of Eq.~(\ref{Eq:dim_result}).
The exact value of $c$, of course, relies on system details such as object shape and chirality,
and it should be a tremendous task to obtain the general expression of $c$ analytically.
However, as we show in the Supplemental Information, a possible upper bound of $c$ can be envisioned, which turns out to be of the order of $10^{-2}$.
In addition, we numerically evaluate $c$ for different objects and flow patterns considered in simulations,
which is indeed found to be small as consistent with the proposed upper bound (see the next section of Simulation results).
Combining these facts, we can take a conservative bound of $|\epsilon|$ for an efficient separation as $|\epsilon| \lesssim {\cal O}(10^{-3})$ for various flow strengths and/or object sizes, even though a rather unrealistically narrow initial distribution ($\delta \sim 1$, i.e., $\sigma_0 \sim \ell$) is assumed. 
Considering the measurement and control accuracy of current microfluidic devices (to our knowledge, of the order of 0.1\% at best),
one might view this range of $\epsilon$ to be synonymous for a singular flow.
It implies that the chiral separation is possible only by quasi-two-dimensional flows described by a velocity gradient tensor $\mathbf{J}$
with a vanishingly small eigenvalue.

Before going further, we explicitly mention the validity range of our theory, especially, in terms of
the relevant object size $\ell$.
Our theoretic formulation assumes the low Reynolds and high P\'{e}clet number conditions,
which give the range of appropriate object size as $(k_BT/\eta V)^{1/3} \ll \ell \ll (\eta/\rho V)^{1/2}$.
Note that $\mathrm{Re} = \rho v \ell /\eta$ where $v\sim V\ell$ and $\rho$ is the fluid density.
For a water at room temperature, $0.01\mu\mathrm{m} \ll \ell \ll 1\mu\mathrm{m}$ for $V \sim 10^6/\mathrm{s}$,
and $0.1\mu\mathrm{m} \ll \ell \ll 100\mu\mathrm{m}$ for $V \sim 10^2/\mathrm{s}$.
The present theory is based on the assumption of linear flow field which is hard to be realized in most cases over extended length and time scales.
For example, the presence of hydrodynamic boundaries arising from confining walls of microfluidic devices leads to nonlinear flows, which
might yield nontrivial effects on the separation.
However, analysis on general nonlinear flow fields is beyond the scope of the present study, and the relevance of our linear flow analysis will be discussed later
in more detail in section of {\bf Possible applications}.

\subsection*{Simulation results}

In order to demonstrate the arguments, 
we perform Langevin dynamics simulations by integrating Eq.~(\ref{Eq:EquationOfMotion}) with explicitly taking into account full hydrodynamic interactions among object elements
at the level of the Rotne-Prager tensor~\cite{Garcia}.
As typical examples of enantiomers, helix and tetrahedral structures are concerned. We decompose the structures into arrays of $N$ closely packed beads with radius $a$. The decomposition allows us to calculate their grand mobility tensors which in turn yield mobility tensors for a rigid body motion of Eq.~(\ref{Eq:EquationOfMotion}), 
according to the conversion equations, Eq.~(19) to (22) in Ref.~\cite{Garcia}
(see the Supplemental Information for details).
In simulations, we rescale all lengths by the bead radius $a$, giving linear size $\ell$ of the helix and tetrahedral structure as $225 a$ and $254 a$, respectively.
Characteristic time scale is $\tau = 6 \pi \eta N a^3/k_B T$.
The distribution of initial positions of objects is assumed to be a Gaussian centered at the origin with an initial width $\sigma_0$.

\begin{figure*}[t]
\center
\includegraphics[width=0.9\linewidth]{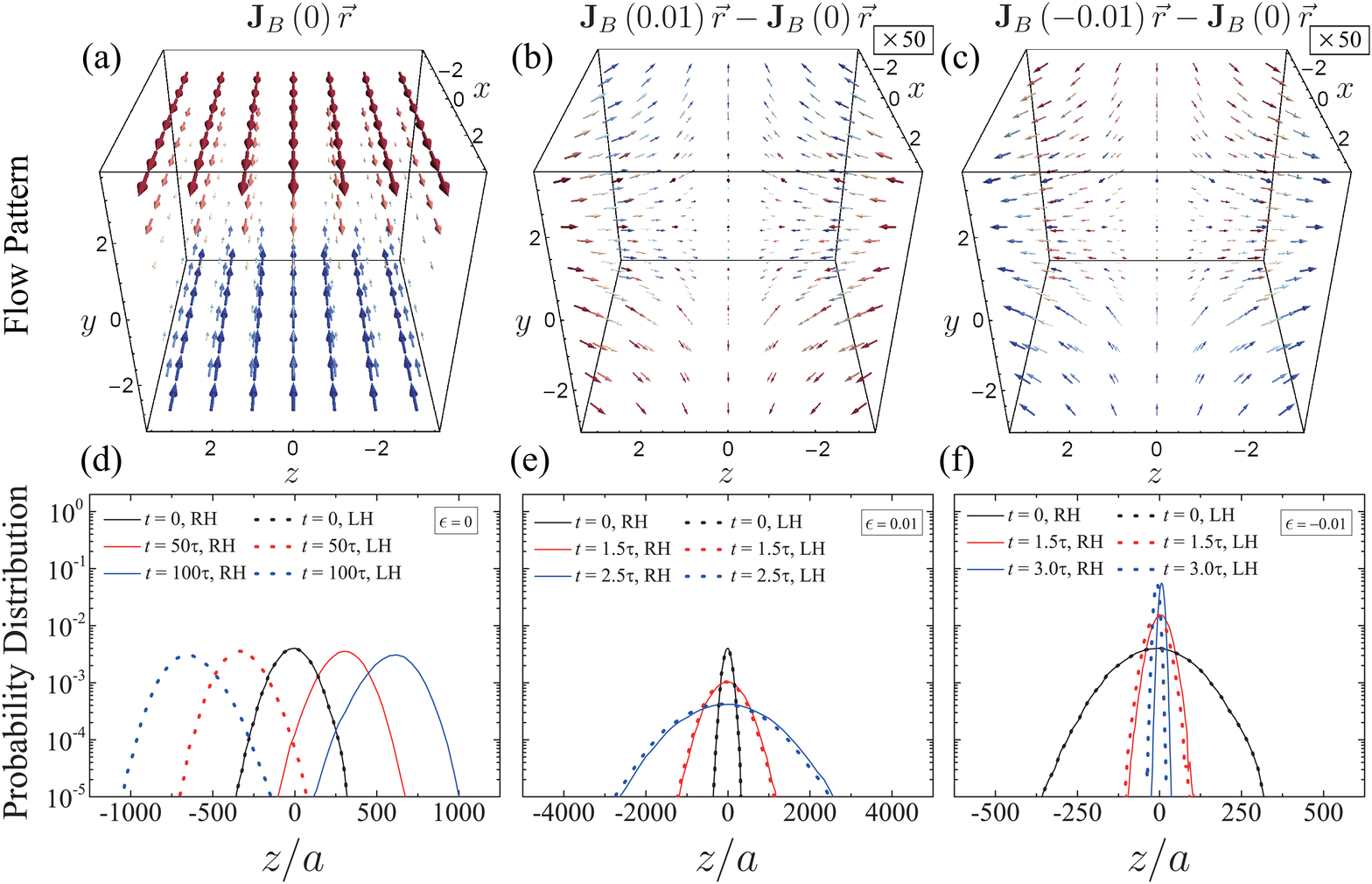}
\caption{
Three-dimensional visualization of flow B with (a) $\epsilon=0$, and difference fields for (b) $\epsilon=0.01$ and (c) $\epsilon=-0.01$. 
Arrow length is proportional to the velocity magnitude, and the color varies from red to blue as $U_x$ changes from positive to negative.
Arrows in (b) and (c) are enlarged by 50 times for visibility. 
(d-f)  Probability distributions of $R$ helix(solid lines) and $L$ helix(dotted lines) are shown at different times for the corresponding flow fields. 
At each case, data are obtained from $10^6$ ensembles of Langevin dynamics simulations with the initial distribution of width, $\sigma_0=100a$.
}
\label{Fig:DistAndPhi}
\end{figure*}

We consider two different types of flow fields, one with a diagonalizable velocity gradient tensor $\mathbf{J}_A$ 
and the other with a non-diagonalizable tensor $\mathbf{J}_B$, parameterized by a dimensionless variable $\epsilon$. 
First, flow-$A$ is described by the velocity gradient tensor,
\begin{eqnarray}
\label{Eq:JA}
{\bf J}_A(\epsilon) &=&
V_A \left(
    \begin{array}{ccc}
    \epsilon & 2 & 0 \\
    1 & -2\epsilon & 0 \\
    0 & 0 & \epsilon~,
    \end{array}
\right)
\end{eqnarray}
which has distinct eigenvalues, $\lambda_{3}/V_A=\epsilon$ and $\lambda_{1,2}/V_A= -(\epsilon \pm \sqrt{8+9\epsilon^{2}})/2$. If $\epsilon=0$, the eigenvalue $\lambda_3$ vanishes and the corresponding ${\bf J}_{A}$ describes a two-dimensional flow.
Flow-$B$ is represented by
\begin{eqnarray}
\label{Eq:JB}
{\bf J}_B(\epsilon) &=&
V_B\left(
    \begin{array}{ccc}
    -\epsilon & 1 & 0 \\
    0 & -\epsilon & 0 \\
    0 & 0 & 2\epsilon
    \end{array}
\right),
\end{eqnarray}
where a finite $\epsilon$ describes a deviation from the shear flow~($\epsilon =0$), and eigenvalues are degenerated, $\lambda_{3}/V_B=2\epsilon$ and $\lambda_{1,2}/V_B=-\epsilon$.
Both flows are incompressible, i.e., ${\bf J}$ is traceless, and 
have a reflection symmetry about $xy$-plane. According to our symmetry argument, the chirality-dependent drift is expected to occur
along the $z$-direction which is the eigenmode direction of $\lambda_{3}$.
Dimensionless flow velocities, $\tilde {V}_i = V_i \tau = 6 \pi \eta N a^3 V_i/k_B T$ for $i=A, B$, are set to 30 in order to realize high Pe;
$\mathrm{Pe} \sim 3.2 \times 10^4$ for helix and $\mathrm{Pe} \sim 3.9 \times 10^4$ for tetrahedral. 
The magnitude of flow velocity gradient, $V$, can be defined in a basis independent way as $V_0=\sqrt{\sum_{i,j}E_{ij}E_{ij}}$. 
However, the current expression of $V$ equals $V_0$ up to a prefactor of the order of one and does not lead to any qualitative difference in results.
For a wide range of flow velocity $V$ and object size $\ell$,
$c$ is found to be small as  $c\lesssim 10^{-3}$ for the helix and $c\lesssim 10^{-2}$ for the tetrahedral,
in accord with the estimate on the possible maximum value of $c$ (see the Supplemental Information).
The previous criterion of Eq.~(\ref{Eq:dim_result}) predicts
that for separation to occur, the upper bound of $|\epsilon|$ is given as $|\epsilon| \lesssim {\cal O} (10^{-3})$
even when a very narrow initial distribution of molecular size is assumed ($\delta \sim 1$).
We test this prediction through numerical simulations, as varying the value of $\epsilon$ of the flows considered above.

Figure~\ref{Fig:DistAndPhi} explicitly shows the flow fields at different flow parameters, $\epsilon = 0$ and $\pm 0.01$, for the flow-$B$ (a-c),
and the time evolutions of corresponding probability distributions of $R$ or $L$ helices, along the $z$-direction, obtained from the simulations (d-f).
Note that, in (b-c), the arrows indicating difference fields are magnified by 50 times for clear visibility.
As consistent with Eq.~(\ref{Eq:dim_result}), a small but finite value of $\epsilon$ leads to qualitatively different behaviors, despite the apparent similarity to the separable shear flow with $\epsilon=0$.
For either small positive or negative $\epsilon$, the probability distributions continue to substantially overlap  even at $t\geq t_\lambda$,
while for $\epsilon=0$, the two distributions are well discriminated and $d$ increases with time,
enabling complete separation.

\begin{figure} [t]
\center
\includegraphics[width=0.9\linewidth]{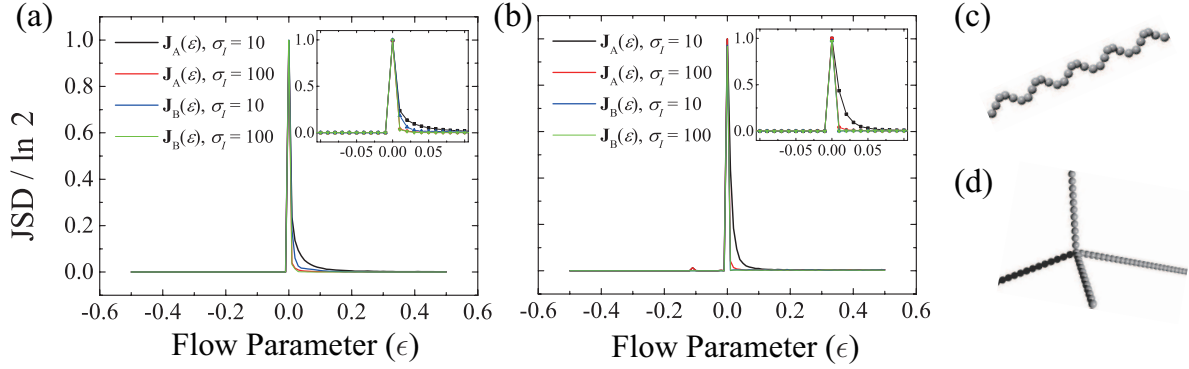}
\caption{
JSD as functions of flow parameter $\epsilon$
for (a) helix and (b) tetrahedral under flow A (solid lines) and B (dashed lines). Two different widths of initial distributions are considered, i.e.,
$\sigma_0 = 10 a$ (black lines) and $100a$ (red lines).
The behaviors near $\epsilon=0$ are magnified in the insets. 
Shown in (c) and (d) are parts of the helix and tetrahedral considered in simulations. 
The helix consists of 25 turns with 6-beads per turn, the radius of $3a$, and the helical pitch of $9a$ ($N=150$). The tetrahedral structure has arm lengths of $200a$, $104a$, $48a$, and $24a$ ($N=189$).
}
\label{Fig:Structure}
\end{figure}

Now we propose to quantify the degree of the separation by the Jensen-Shannon divergence (JSD):
\begin{equation}\label{jsd}
\mbox{JSD}(p^{L}||p^{R})=\frac{1}{2}\sum_{\alpha} \int dz \, p^{\alpha}(z)\ln \left[\frac{2p^{\alpha}(z)}{p^{L}(z)+p^{R}(z)}\right], \nonumber
\end{equation}
where $p^{\alpha}(z)$ represents the probability distribution to find an object with chirality $\alpha=L,R$ 
at a projected position $z$ along a chosen axis of observation.
Unlike $d/\sigma$, JSD is bounded as $0\leq \mbox{JSD} \leq \ln 2$, and its value depends on overlapping area between $p^{L}(z)$ and $p^{R}(z)$.
If $p^{L}(z)=p^{R}(z)$ (perfect overlap), JSD vanishes.
If $p^{L}(z)$ has no overlapping region with $p^{R}(z)$ (complete separation), JSD reaches its maximum value $\ln 2$. 
Any values of JSD less than $\ln 2$ signal that finite overlap between $p^{L}(z)$ and $p^{R}(z)$ exists, and resulting separation is inaccurate.  
JSD therefore provides the well-defined scale of separation precision, taking into account shape details of $p^{\alpha}(z)$.

Depicted in Fig.~\ref{Fig:Structure} is JSD as a function of $\epsilon$ for different flow patterns and chiral objects.
It follows from Eq.~(\ref{Eq:shorttimes}) and (\ref{Eq:longtimes}) that when $\lambda=0$, $d/\sigma > 1$ for $t > t_s$,
while for $\lambda \neq 0$, it is saturated to a constant value after $t_\lambda$.
The occurrence of macroscopic separation can therefore be determined by measuring JSD at time $t$ larger than $t_s $ (for $\lambda=0$) and $t_\lambda$ (for $\lambda\neq0$)
which are estimated as follows.
For the high Pe regimes considered in this work, $t_s$ is given by $t_\sigma \sim \delta/c V$.
In simulations, the upper bound of $\delta$ is ${\cal O} (1)$, the lower bound of $c$ is ${\cal O}(10^{-4})$, and $V= \tilde{V}/\tau = 30/\tau$,
which leads to $t_s \sim {\cal O}(10^2 \tau)$.
For a non-zero $\lambda$, $t_\lambda = 1/|\lambda| \sim 1/|\epsilon| V \sim \tau/|\epsilon| \tilde{V}$,
and the smallest value of $|\epsilon|$ other than zero is $0.01$ in Fig.~\ref{Fig:Structure}.
Thus, JSD's are measured at $t \sim {\cal O}(10^3 \tau)$ for $\epsilon \leq 0$, ${\cal O}(10 \tau)$ for $0 < \epsilon \leq 0.2$, and ${\cal O}(\tau)$ for $\epsilon > 0.2 $,
which are long enough for JSD to reach its stationary value.
For numerical evaluations of JSD, the probability distributions $p(z)$ are discretized into histograms with the bin size of the order of the object size.
The complete separation with the maximum JSD of $\ln 2$ is achieved in all considered cases only when $\epsilon$ (and thus at least one of the eigenvalues) is vanishingly small.
A finite value of JSD results for small positive $\epsilon$ 
from unrealistically narrow initial dispersions of particles ($\sigma_0/\ell \sim 0.05$).
In most realizable situations where $\sigma_0 \gg \ell$, it is more pronounced that JSD yields very small values unless $\epsilon$ is vanishingly small.
As shown in Fig.~\ref{Fig:Structure}, the simulation results clearly demonstrate our claim that the complete 
separation indeed occurs by quasi-two-dimensional flows satisfying the condition, Eq.~(\ref{Eq:dim_result}).

\begin{figure*}
	\center
	\includegraphics[width=0.9\linewidth]{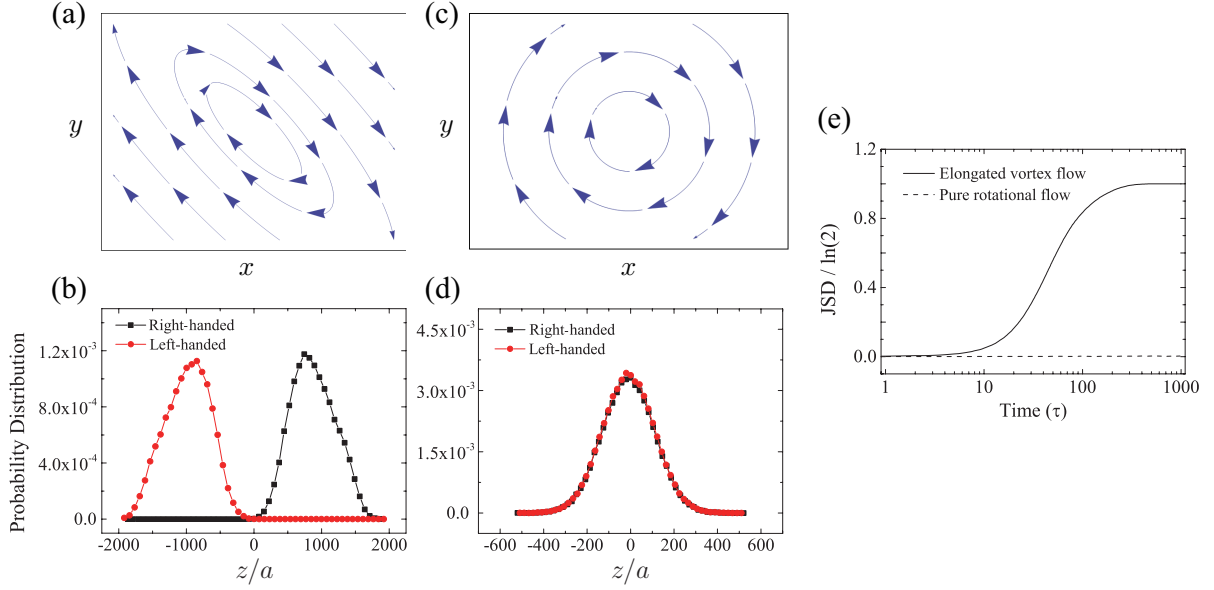}
	\caption{ 
	Two-dimensional flow streamlines for (a) an elongated vortex flow (Eq.~(\ref{Eq:elongated})) and (c) a pure rotational flow (Eq.~(\ref{Eq:rotation})) in the $xy$-plane. Probability distributions of right-handed (black) and left-handed (red) helices, along the $z$-direction (in units of $a$), at $t=500\tau$ under (b) the elongated vortex flow and (d) the pure rotational flow. (e) JSDs as a function of time (in units of $\tau$) for the respective flow patterns. The chiral separation is clearly achieved for the elongated vortex flow, but not for the pure rotational flow. Initial distributions are given as Gaussian centered at the origin with dispersion of $100 a$, the number of ensembles is $10^5$, and the same helical objects are considered as in Fig.~\ref{Fig:Structure}. $\tilde{V}=30$.
	}
	\label{Fig:Vortex}
\end{figure*}

\subsection*{Possible applications}
	
Finally, we discuss the relevance of our linear flow analysis in terms of possible practical applications. 
In most cases, it is challenging to realize linear flows persisting for extended length and time scales in microfluidic devices.
A simple shear flow is a well-known exception.
We exhibit here two other examples where the linearity of flows is easily assured and at the same time,
the present analysis can be useful for practical purposes.
In particular, we show that for chiral separations, it is enough to have a linear flow only in a localized region
if a confining potential is applied together.

\subsubsection*{Vortex flow} \label{Sec:Vortex} 

According to our theoretic formulations,
the chiral separation occurs when both of the following conditions have to be satisfied:
First, one of the eigenvalues of velocity gradient tensor should be much smaller than the inverse of the separation time scale (the eigenmode analysis).
Secondly, there should exist a non-vanishing rate-of-strain field that induces a finite drift velocity (the parity-inversion argument).
Now we present a salient example explicitly showing the indispensable role of the rate-of-strain field:
consider a pure rotational flow with a perfect circular streamline, defined by the velocity gradient tensor, 
\begin{equation}
\label{Eq:rotation}
\mathbf{J} = V
 \left(
\begin{array} {ccc}
0 & 1 & 0 \\
-1 & 0 & 0 \\
0 & 0 & 0
\end{array}
\right)~.
\end{equation}
This flow satisfies the first condition but obviously not the second condition because it has a vanishing rate-of-strain field, $\mathbf{E} =0$.
Hence it cannot separate any kind of chiral pairs.
On the contrary, a vortex flow deformed by a finite rate-of-strain field, e.g., with the following velocity gradient tensor,
\begin{equation}
\label{Eq:elongated}
\mathbf{J} = V
 \left(
\begin{array} {ccc}
0.8 & 1 & 0 \\
-1 & -0.8 & 0 \\
0 & 0 & 0 
\end{array}
\right)~
\end{equation}
fulfills the both conditions and could, therefore, lead to the separations, as indeed confirmed by the Langevin dynamics simulations (Fig.~\ref{Fig:Vortex}).
The vortex flow can be an obvious solution for chiral separations, if not perfectly circular, 
and is another example of linear flows that persist for an extended period of time in microfluidic setups.
As shown here, the vortex flows with circular streamlines have very distinct separation powers, depending on whether or not they have
a finite rate-of-strain field. 

Our prediction on the separation power of vortex flows can be tested by a microfluidic four-roll mill device suggested by Lee et. al. which can produce the entire spectrum of flow types, from purely rotational flow to purely extensional flow, by varying flow rate ratio~\cite{Muller}.
Considering the dimensions of the device, the sub-micron helical objects ($\ell \lesssim \mu$m), for example, should be observed to be separated
for elongated vortex flow but not for purely rotational flow. Also using various chiral objects of different shapes in this setup can be a feasible way to confirm our theory predicting that the flow properties rather than object shapes are essential to chiral separation.

\begin{figure}
	\center
	\includegraphics[width=0.9\linewidth]{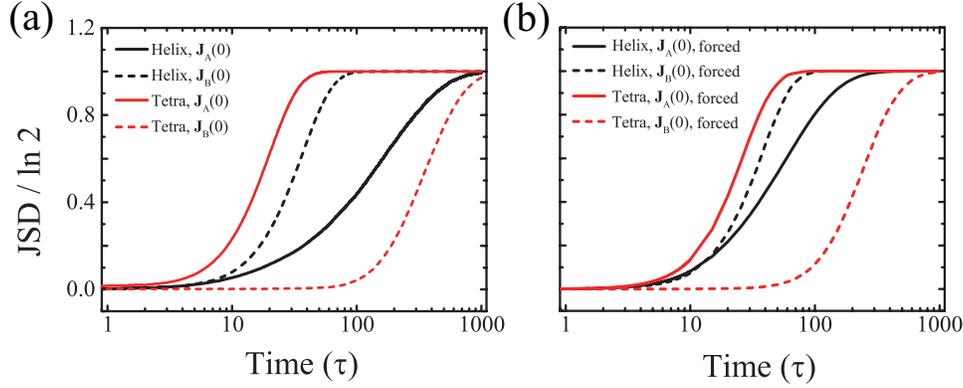}
	\caption{Temporal evolution of JSD for helix (black) and tetrahedral (red) under flow A (solid lines) and flow B (dashed lines). Here, we set $\epsilon=0$, $\sigma_0 = 100a$, $\tilde{V}=30$, 
	and time is in units of $\tau$. 
	(Left panel) In the absence of external potential, JSD reaches its maximum of $\ln 2$, indicating a complete chiral separation, for all considered cases around $t \sim {\cal O}(10^3) \tau$. 
	(Right panel) In the presence of confining potential $\phi(x,y) = k(x^2 + y^2)$ with $k=8 k_B T/a^2$, JSD shows the very similar behaviors and the complete separation is also achieved for all cases.
	For helix and tetrahedral, the same objects as in Fig.~\ref{Fig:Structure} are considered, the number of ensembles is $10^5$, and 
	the initial distributions are given as Gaussian centered at the origin with dispersion of $100 a$.
	}
	\label{Fig:Forced}
\end{figure}

\subsubsection*{Confining potential} \label{Sec:Confining}

At low Reynolds numbers, the resistance formalism linearly relates the hydrodynamic net force and torque
exerted on a rigid object to the flow parameters~\cite{Sangtae}.
Due to the linearity of the relations, originating from the linearity of the Stokes equations, 
the external forces such as confining force can be separately added into the equations of motion (see the Supplemental Information).
Consequently, even in the presence of external forces our analysis can be performed in a similar way and the separation criterion remains intact in general.

In order to numerically verify this, we perform Langevin dynamics simulations with Flow-A and Flow-B previously defined, 
considering helical and tetrahedral objects under an external potential, $\phi(x, y) = k (x^2 + y^2)$. 
Here, we set $\epsilon=0$, and then Flow-A and B represent a two-dimensional extensional flow and a shear flow, respectively.
The external potential plays a role as a two-dimensional confinement, constraining particle positions near to the origin in the $xy$ plane.
We note that for both flows, the separation occurs along the $z$-axis.
The right panel of Fig.~\ref{Fig:Forced} shows the simulation results of temporal evolution of JSD in the presence of the potential
while the left panel exhibits the results without the potential.
As deduced, one can see that the separation behaviors remains the same qualitatively.
This suggests that for practical applications of our analysis, a linear flow field does not necessarily have to persist over extended length and time scales.
The two-dimensional confinement potential makes the separation (e.g., along the $z$-axis) take place only in a limited space of the $xy$-plane where
the linearity of flows can be rather easily assumed for a time longer than the separation time scale.

\section*{Discussion}

In practical applications of harnessing microfluidic devices for separations, one of the central questions will be to determine which flow has separation capability.
Despite increasing attention to microfluidic chiral resolutions, most of the previous works are restricted to be considering an object of a specific shape in a given flow pattern.
The complicated mathematical structure of the equations of motion present difficulties to understand chiral separation phenomena even for a specific flow, and 
a comprehensive picture of common mechanisms and general conditions for chiral separations in terms of flow properties has been lacking. 
We have tackled this problem for an arbitrary linear flow using simple ideas, namely, considering symmetry properties and adopting the eigenmode analysis of flow fields.
This enables us to draw an intuitive physical picture of the underlying mechanism of chiral separation dynamics.
According to our results, the common features of separable flows are summarized as i) flows with a finite strain-rate tensor
and ii) quasi-two-dimensional flows with small eigenvalues obeying Eq.~(\ref{Eq:criterion}).
The typical examples satisfying both conditions are shear flow, vortex flow, and two-dimensional extensional flow,
all of which are indeed demonstrated here to cause separations via the Langevin dynamics simulations.
The present study thus provides a theoretical understanding of why two-dimensional flows such as shear and vortex flow are efficient to induce chiral separations.
Our results provide simple criteria that would allow us to categorize and decide what kind of flows have a separation power or not.
This is the prediction that cannot be easily made, without complicated numerical calculations or extensive simulations, from the theoretical studies known hitherto.
It is also important to note that the separation criteria only concern the properties of flows, not of objects,
so that they are applicable for objects of different shapes.

\section*{Acknowledgements}
This research was supported by Basic Science Research Program through the National Research Foundation of Korea(NRF) funded by the Ministry
of Education, Science and Technology(Grant No. NRF-2013R1A1A2013137).

\section*{Author Contributions}
Y.W.K. supervised the project, S.R. carried out the calculations under the guidance of J.Y. and Y.W.K. All authors discussed the results, wrote and reviewed the manuscript.

\section*{Additional Information}
{\bf Competing financial interests:} The authors declare no competing financial interests.


\begin{thebibliography}{10}

\bibitem{Stalcup} Stalcup,~A.~M. Chiral separations. {\it Annu. Rev. Anal. Chem.} {\bf 3}, 341 (2010).

\bibitem{Ariens} Ari\"{e}ns,~E.~J. Stereochemistry, A basis for sophisticated nonsense in pharmacokinetics and clinical pharmacology. {\it Eur. J. Clin. Pharmacol.} {\bf 26}, 663 (1984).

\bibitem{Maier} Maier,~N.~M., Fronco,~P. \& Lindner,~W. Separation of enantiomers: needs, challenges, perspectives. {\it J. Chromatogr. A.} {\bf 906}, 3 (2001).

\bibitem{deGennes} De Gennes,~P.~G. Mechanical selection of chiral crystals. {\it Europhys. Lett.} {\bf 46}, 827 (1999).

\bibitem{Baranova} Baranova,~N.~B. \& Zel'dovich,~B.~Y. Separation of mirror isomeric molecules by radio-frequency electric field of rotating polarization. {\it Chem. Phys. Lett.} {\bf 57}, 435 (1978).

\bibitem{Schamel} Schamel,~D., Pfeifer,~M., Gibbs,~J.~G., Miksch,~B., Mark,~A.~G. \& Fischer,~P. Chiral colloidal molecules and observation of the propeller effect. {\it J. Am. Chem. Soc.} {\bf 135}, 12353 (2013).

\bibitem{Howard} Howard,~D.~W., Lightfoot,~E.~N. \& Hirschfelder,~J.~O. Hydrodynamic resolution of optical isomers. {\it AIChe J.} {\bf 22}, 794 (1976).

\bibitem{Kim} Kim,~Y.~J. \& Rae,~W.~J. Separation of screw-sensed particles in a homogeneous shear field. {\it Int. J. Multiphas. Flow.} {\bf 17}, 717 (1991).

\bibitem{Chen1} Chen,~P. \& Chao,~C.~H. Lift forces of screws in shear flows. {\it Phys. Fluids} {\bf 19}, 017108 (2007).

\bibitem{Makino} Makino,~M., Arai,~L. \& Doi,~M. Shear migration of chiral particle in parallel-disk. {\it J. Phys. Soc. Jpn.} {\bf 77}, 064404 (2008).

\bibitem{Marcos} Marcos, Fu,~H.~C., Powers,~T.~R. \& Stocker,~R. Separation of microscale chiral objects by shear flow. {\it Phys. Rev. Lett.} {\bf 102}, 158103 (2009).

\bibitem{Maria} Aristov,~M., Eichhorn,~R. \& Bechinger,~C. Separation of chiral colloidal particles in a helical flow field. {\it Soft Matter} {\bf 9}, 2525 (2013).

\bibitem{Hermans} Hermans,~T.~M., Bishop,~K.~J.~M., Stewart,~P.~S., Davis,~S.~H. \& Grzybowski,~B.~A. Vortex flows impact chirality-specific lift forces. {\it Nat. Commun.} {\bf 6}, 5640 (2015).

\bibitem{Kostur} Kostur,~M., Schindler,~M., Talkner,~P. \& H\"{a}nggi,~P. Chiral separation in microflows. {\it Phys. Rev. Lett.} {\bf 96}, 014502 (2006).

\bibitem{Meinhardt} Meinhardt,~S., Smiatek,~J., Eichhorn,~R. \& Schmid,~F. Separation of chiral particles in micro- or nanofluidic channels. {\it Phys. Rev. Lett.} {\bf 108}, 214504 (2012).

\bibitem{Eichhorn} Eichhorn,~R. Microfluidic sorting of stereoisomers. {\it Phys. Rev. Lett.} {\bf 105}, 034502 (2010).

\bibitem{Chen} Chen,~P. \& Zhang,~Q.~Y. Dynamical solutions for migration of chiral DNA-type objects in shear flows. {\it Phys. Rev. E} {\bf 84}, 056309 (2011).

\bibitem{Watari} Watari,~N. \& Larson,~R.~G. Shear-induced chiral migration of particles with anisotropic rigidity. {\it Phys. Rev. Lett.} {\bf 102}, 246001 (2009).

\bibitem{Talkner} Talkner,~P., Ingold,~G.~L. \& H\"{a}nggi,~P. Transport of flexible chiral objects in a uniform shear flow. {\it New. J. Phys.} {\bf 14}, 073006 (2012).

\bibitem{Happel} Happel,~J. \& Brenner,~H. {\it Low Reynolds Number Hydrodynamics} (Martinus Nijhoff Publishers, Hague, 1983).

\bibitem{Sangtae} Kim,~S. \& Karrila,~S.~J. {\it Microhydrodynamics: Principles and Selected Applications} (Butterworth-Heinemann, Boston, 1991).

\bibitem{Brenner} Brenner,~H. The stokes resistance of an arbitrary particle-III. {\it Chem. Eng. Sci.} {\bf 19}, 631 (1964).

\bibitem{Garcia} Carrasco,~B. \& Garcia de la Torre,~J. Improved hydrodynamic interaction in macromolecular bead models. {\it J. Chem. Phys.} {\bf 111}, 4817 (1999).



\bibitem{Muller} Lee,~J.~S., Dylla-Spears,~R., Teclemariam,~N.~P. \& Muller,~S.~J. Microfluidic four-roll mill for all flow types, Appl. Phys. Lett. {\bf 90} 074103 (2007).





\end{thebibliography}
\end{document}